\documentclass[iop,numberedappendix,appendixfloats]{emulateapj}
\usepackage{times}
\usepackage{appendix}

\begin{document}

\newcommand{\mirlum}{L_{\rm 8}}
\newcommand{\ebmvstars}{E(B-V)_{\rm stars}}
\newcommand{\ebmvgas}{E(B-V)_{\rm neb}}
\newcommand{\ebmv}{E(B-V)}
\newcommand{\lha}{L(H\alpha)}
\newcommand{\lir}{L_{\rm IR}}
\newcommand{\lbol}{L_{\rm bol}}
\newcommand{\luv}{L_{\rm UV}}
\newcommand{\rs}{{\cal R}}
\newcommand{\ugr}{U_{\rm n}G\rs}
\newcommand{\ks}{K_{\rm s}}
\newcommand{\gmr}{G-\rs}
\newcommand{\hi}{\text{\ion{H}{1}}}
\newcommand{\nhi}{N(\text{\ion{H}{1}})}
\newcommand{\lognhi}{\log[\nhi/{\rm cm}^{-2}]}
\newcommand{\molh}{\text{H}_2}
\newcommand{\nmolh}{N(\molh)}
\newcommand{\lognmolh}{\log[\nmolh/{\rm cm}^{-2}]}
\newcommand{\oii}{\text{[\ion{O}{2}]}}
\newcommand{\neiii}{\text{[\ion{Ne}{3}]}}
\newcommand{\oiii}{\text{[\ion{O}{3}]}}
\newcommand{\nii}{\text{[\ion{N}{2}]}}
\newcommand{\hei}{\text{\ion{He}{1}}}
\newcommand{\ha}{\text{H$\alpha$}}
\newcommand{\hb}{\text{H$\beta$}}
\newcommand{\hg}{\text{H$\gamma$}}
\newcommand{\hd}{\text{H$\delta$}}
\newcommand{\he}{\text{H$\epsilon$}}
\newcommand{\hz}{\text{H$\zeta$}}
\newcommand{\hn}{\text{H$\eta$}}
\newcommand{\htheta}{\text{H$\theta$}}
\newcommand{\hiota}{\text{H$\iota$}}
\newcommand{\hk}{\text{H$\kappa$}}
\newcommand{\caii}{\text{\ion{Ca}{2}}}
\newcommand{\sii}{\text{[\ion{S}{2}]}}
\newcommand{\delw}{\Delta\log[W/{\rm \AA}]}
\newcommand{\delsfr}{\Delta\log[{\rm SFR(\ha)}/M_\odot\,{\rm yr}^{-1}]}
\newcommand{\delz}{\Delta[12+\log({\rm O/H})_{\rm N2}]}
\newcommand{\delion}{\Delta\log({\rm O32})}
\newcommand{\sfrha}{\text{SFR[H$\alpha$]}}
\newcommand{\sfrsed}{\text{SFR[SED]}}
\newcommand{\ssfrha}{\text{sSFR[H$\alpha$]}}
\newcommand{\ssfrsed}{\text{sSFR[SED]}}

\title{The MOSDEF Survey: The First Direct Measurements of the Nebular Dust Attenuation Curve at High Redshift\altaffilmark{*}}

\author{\sc Naveen A. Reddy\altaffilmark{1},
Alice E. Shapley\altaffilmark{2},
Mariska Kriek\altaffilmark{3},
Charles C. Steidel\altaffilmark{4},
Irene Shivaei\altaffilmark{5,6},
Ryan L. Sanders\altaffilmark{6,7}, 
Bahram Mobasher\altaffilmark{1},
Alison L. Coil\altaffilmark{8},
Brian Siana\altaffilmark{1},
William R. Freeman\altaffilmark{1},
Mojegan Azadi\altaffilmark{9},
Tara Fetherolf\altaffilmark{1},
Gene Leung\altaffilmark{8},
Sedona H. Price\altaffilmark{10},
Tom Zick\altaffilmark{3}}

\altaffiltext{1}{Department of Physics and Astronomy, University of California, 
Riverside, 900 University Avenue, Riverside, CA 92521, USA; naveenr@ucr.edu}
\altaffiltext{2}{Department of Physics \& Astronomy, University of California,
Los Angeles, 430 Portola Plaza, Los Angeles, CA 90095, USA}
\altaffiltext{3}{Astronomy Department, University of California, Berkeley,
Berkeley, CA 94720, USA}
\altaffiltext{4}{Cahill Center for Astronomy and Astrophysics, California Institute of Technology, MC 249-17, Pasadena, CA 91125, USA}
\altaffiltext{5}{Steward Observatory, University of Arizona, 933 North 
Cherry Avenue, Tucson, AZ 85721, USA}
\altaffiltext{6}{NASA Hubble Fellow}
\altaffiltext{7}{Department of Physics, University of California, Davis, One Shields Ave, Davis, CA 95616, USA}
\altaffiltext{8}{Center for Astrophysics and Space Sciences, University of
California, San Diego, 9500 Gilman Drive, La Jolla, CA 92093-0424, USA}
\altaffiltext{9}{Harvard-Smithsonian Center for Astrophysics, 60 Garden Street, Cambridge, MA 02138, USA}
\altaffiltext{10}{Max-Planck-Institut f\"{u}r Extraterrestrische Physik, Postfach 1312, Garching, D-85741, Germany}

\altaffiltext{*}{Based on data obtained at the W.M. Keck Observatory,
  which is operated as a scientific partnership among the California
  Institute of Technology, the University of California, and NASA, and
  was made possible by the generous financial support of the W.M. Keck
  Foundation.}

\slugcomment{DRAFT: \today}

\begin{abstract}

We use a sample of 532 star-forming galaxies at redshifts $z\simeq
1.4-2.6$ with deep rest-frame optical spectra from the MOSFIRE Deep
Evolution Field (MOSDEF) survey to place the first constraints on the
nebular attenuation curve at high redshift.  Based on the first five
low-order Balmer emission lines detected in the composite spectra of
these galaxies ($\ha$ through $\he$), we derive a nebular attenuation
curve that is similar in shape to that of the Galactic extinction
curve, suggesting that the dust covering fraction and
absorption/scattering properties along the lines-of-sight to massive
stars at high redshift are similar to those of the average Milky Way
sightline.  The curve derived here implies nebular reddening values
that are on average systematically larger than those derived for the
stellar continuum.  In the context of stellar population synthesis
models that include the effects of stellar multiplicity, the
difference in reddening of the nebular lines and stellar continuum may
imply molecular cloud crossing timescales that are a factor of $\ga
3\times$ longer than those inferred for local molecular clouds,
star-formation rates that are constant or increasing with time such that
newly-formed and dustier OB associations always dominate the ionizing
flux, and/or that the dust responsible for reddening the nebular
emission may be associated with non-molecular (i.e., ionized and
neutral) phases of the ISM.  Our analysis points to a variety of
investigations of the nebular attenuation curve that will be enabled
with the next generation of ground- and space-based facilities.

\end{abstract}


\section{\bf INTRODUCTION}
\label{sec:intro}

Recent advances in near-infrared detector technology and multiplexing
capabilities have led to a rapid increase in the number of rest-frame
optical ($\lambda \simeq 3700-6700$\,\AA) line measurements for
high-redshift ($1.4\la z\la 3.8$) galaxies, numbering now in the
thousands (e.g., \citealt{forster09, kashino13, steidel14, kriek15}).
In turn, these measurements have yielded valuable insights into the
dust reddening, gas-phase metallicities, star-formation rates (SFRs)
and physical state of the gas in the ISM of high-redshift galaxies.
Critical to many of these inferences is the wavelength dependence of
dust obscuration of the ionized gas (i.e., the nebular dust
attenuation curve), typically assumed to follow that of the average
Milky Way sightline \citep{cardelli89}.

The customary approach to deducing the attenuation of the nebular
emission and stellar continuum in galaxies is to assume that each is
subject to a different attenuation curve (even if the intrinsic dust
{\em extinction} curve is the same throughout the galaxy), stemming
from the expectation that the young stars that dominate the nebular
emission are located preferentially in regions with higher dust
covering fractions associated with their parent molecular clouds
(e.g., with a geometry approximating a foreground screen of dust;
\citealt{calzetti94}).  In particular, for nearby starburst galaxies,
the Galactic extinction curve \citep{cardelli89} is commonly adopted
for the nebular line emission, whereas an {\em attenuation} curve
(i.e., one that accounts for the scattering of light into the
line-of-sight and for a non-uniform distribution of column densities;
e.g., \citealt{calzetti00}) is assumed for the stellar continuum
(e.g., \citealt{calzetti94, calzetti97}).  Separately, a number of
studies of both local \citep{fanelli88, calzetti97, calzetti00,
  wild11, kreckel13} and high-redshift star-forming galaxies (e.g.,
\citealt{forster09, yoshikawa10, wuyts11, kashino13, wuyts13, price14,
  reddy15, debarros16, buat18, shivaei20}) have found that the line
emission arising from the nebular (ionized gas) regions is subject to
a higher degree of reddening than the non-ionizing stellar continuum
emission, perhaps reflecting a variation in the column density of dust
as viewed along the average sightlines to differently-aged stellar
populations within galaxies.  Together, the variations in the shape of
the attenuation curve and column density of dust point to a
complicated geometry of dust and stars that varies significantly from
galaxy to galaxy.

There has been much debate regarding the appropriate curve to use for
the stellar continuum in high-redshift galaxies.  However, the shape
of the nebular attenuation curve---and whether it is similar to that
of the stellar continuum---has received little attention despite its
importance in deducing several key physical attributes of the ISM,
such as gas-phase metallicity and ionization state.  Given that the
nebular attenuation curve underpins many of the most fundamental
properties that we can infer for the ISM in galaxies, and that there
is no {\em a priori} reason why the curve should follow that of the
average Milky Way sightline (e.g., due to potential differences in
grain composition and/or distribution, structure of molecular clouds,
etc.), it is imperative to directly probe the shape and normalization
of this curve at high redshift.  From a practical standpoint, the
derivation of the nebular attenuation curve is simpler than that of
the stellar attenuation curve: the former can be accomplished using
recombination emission lines whose strengths are dictated by
well-understood physics, while the latter typically relies on
uncertain assumptions for the intrinsic stellar spectrum of a galaxy.
At any rate, comparison of the nebular and stellar attenuation curves
provides important insights into the spatial distribution and
properties of the dust for differently-aged stellar populations in
galaxies.

Only with the recent accumulation of a large number of spectroscopic
measurements of recombination emission lines in high-redshift galaxies
has it been possible to directly constrain both the shape and
normalization of the nebular dust attenuation curve.  Here, we use
extensive spectroscopy of the Balmer recombination emission lines of
high-redshift ($z\simeq 1.4-2.6$) star-forming galaxies in the MOSFIRE
Deep Evolution Field (MOSDEF; \citealt{kriek15}) survey to place the
first direct constraints on the nebular attenuation curve at high
redshift.

The translation between a single Balmer emission line ratio (e.g.,
$\ha/\hb$) and the nebular reddening, $\ebmvgas$, depends on the shape
of the extinction/attenuation curve.  Consequently, a single line
ratio cannot be used to uniquely identify the shape of the dust curve.
However, with multiple Balmer emission line ratios constructed with at
least three Balmer recombination lines, we can simultaneously
constrain both the reddening and the shape of the dust curve.  The
deep near-IR spectroscopy of the MOSDEF survey covers different
subsets of the Balmer recombination emission lines, depending on the
redshift of each galaxy.  For this study, we have focused on the first
five low-order Balmer emission lines listed in
Table~\ref{tab:lineproperties} that are significantly detected in
either individual galaxy spectra ($\ha$, $\hb$, and occasionally $\hg$
and $\hd$) or in composite spectra of ensembles of galaxies.

The outline of this paper is as follows.  Section~\ref{sec:sample}
describes the parent MOSDEF sample, and some of the steps in the
spectroscopic data reduction that are salient to the determination of
the nebular attenuation curve.  We also describe the method used for
constructing composite spectra and measuring line ratios from these
composites.  Section~\ref{sec:sampleconstruction} focuses on the
various subsamples used to construct composites, while the calculation
of the shape of the nebular attenuation curve is presented in
Section~\ref{sec:calculations}.  In Section~\ref{sec:results}, we
compare the nebular attenuation curve with other common
extinction/attenuation curves.  In Section~\ref{sec:discussion}, we
present a comparison of nebular and stellar reddening, and discuss
differences in these two quantities in the context of stellar
population models that include the effects of stellar multiplicity.
Wavelengths are presented in the vacuum frame.  We adopt a cosmology
with $H_{0}=70$\,km\,s$^{-1}$\,Mpc$^{-1}$, $\Omega_{\Lambda}=0.7$, and
$\Omega_{\rm m}=0.3$.

\begin{deluxetable}{lccc}
\tabletypesize{\footnotesize}
\tablewidth{0pc}
\tablecaption{Balmer Recombination Lines}
\tablehead{
\colhead{Line} &
\colhead{$\lambda$ (\AA)\tablenotemark{a}} & 
\colhead{$I$\tablenotemark{b}} & 
\colhead{Fitting Window (\AA)\tablenotemark{c}}}
\startdata
\ha\tablenotemark{d} & 6564.60 & 2.860 & 6482 - 6652 \\
\hb & 4862.71 & 1.000 & 4813 - 4913 \\
\hg & 4341.69 & 0.468 & 4301 - 4383 \\
\hd & 4102.89 & 0.259 & 4055 - 4160 \\
\he\tablenotemark{e} & 3971.20 & 0.159 & 3946 - 4024 
\enddata
\tablenotetext{a}{Rest-frame vacuum wavelength, taken from the Atomic Spectra Database website of
the National Institute of Standards and Technology (NIST), https://www.nist.gov/pml/atomic-spectra-database.}
\tablenotetext{b}{Intensity of line relative to $\hb$ for Case B recombination, $T_{\rm e}=10000$\,K,
and $n_{\rm e}=100$\,cm$^{-3}$.}
\tablenotetext{c}{Wavelength window over which line fitting was performed.}
\tablenotetext{d}{$\ha$ was fit simultaneously with the $\nii$ doublet (Section~\ref{sec:ratiomeasurements}).}
\tablenotetext{e}{$\he$ is blended with $\neiii\lambda 3969$.  The contribution of the latter
was estimated by measuring $\neiii\lambda 3870$ (Section~\ref{sec:sampleconstruction}), 
for which we assumed a
rest-frame vacuum wavelength of $3869.86$\,\AA\, and a fitting window that spans the range
$\lambda = 3851 - 3882$\,\AA.}
\label{tab:lineproperties}
\end{deluxetable}

\section{\bf SURVEY AND BASIC MEASUREMENTS}
\label{sec:sample}

\subsection{MOSDEF Survey}

The MOSDEF survey \citep{kriek15} used the MOSFIRE instrument
\citep{mclean12} on the Keck telescope to acquire moderate resolution
($R\sim 3000-3600$) rest-frame optical spectra of $\approx 1500$ {\em
  H}-band selected galaxies at redshifts $1.4\la z\la 3.8$ in the
CANDELS fields \citep{grogin11, koekemoer11}.  Galaxies were targeted
for spectroscopy based on pre-existing spectroscopic, grism, or
photometric redshifts that placed them in three redshift ranges ---
$z=1.37-1.70$, $z=2.09-2.70$, and $z=2.95-3.80$ --- where the strong
rest-frame optical emission lines lie in the {\em YJH}, {\em JHK}, and
{\em HK} transmission windows, respectively.  Spectral data were
reduced and extracted as described in \citet{kriek15}.

\subsection{Slit Loss Corrections}
\label{sec:slitlosscorrections}

The method employed here to determine the nebular attenuation curve
relies on taking ratios of multiple Balmer emission lines.  In general,
these lines fall in multiple near-IR filters that were observed in different
weather and seeing conditions.
Reliable determinations of the line
ratios thus require accurate relative flux calibration of the spectra
taken in different filters.  To this end, bright ``slit stars'' were
observed simultaneously with the target galaxies in order to flux
calibrate the spectra and compute first-order corrections for slit
losses.  Because the galaxies are spatially resolved given their
typical sizes and the seeing of the observations, second-order
corrections for slit losses---based on modeling the light profiles of
the galaxies---were applied to the spectra \citep{kriek15, reddy15}.
The efficacy of our slit loss correction procedure was evaluated by
comparing the spectroscopic flux densities of galaxies detected in the
continuum with their broadband flux densities.  This comparison shows
that the slit-loss-corrected spectra yield flux densities that are
typically within $\simeq 18\%$ of the broadband values (e.g.,
\citealt{reddy15}).  Moreover, averaging the slit-loss corrected
spectra for individual galaxies results in composite spectra
whose shapes agree with those of the average broad-band SEDs of the
same galaxies (Section~\ref{sec:composites}).

\subsection{Line Flux Measurements}

Line fluxes for individual objects were measured from the spectra by
fitting Gaussian functions on top of a linear continuum.  Two Gaussian
functions were used to fit the $\oii$ doublet, while three were used
to simultaneously fit $\ha$ and the $\nii$ doublet.  Gaussian
functions were assumed for all other lines.  As the continuum is
generally not detected in the spectra of individual galaxies, the
fluxes of Balmer emission lines were also calculated assuming an
underlying linear continuum.  As such, the Balmer emission-line
measurements for individual galaxies are not corrected for underlying
Balmer absorption.  However, as we discuss below, Balmer absorption
{\em is} taken into account when measuring average line fluxes in the
composite spectra of galaxies.  Errors in line fluxes were derived by
allowing the spectra to vary 500 times according to the error spectra,
and remeasuring the line fluxes from these realizations.  For this
study, AGNs were excluded based on the IR, X-ray, and rest-frame
optical line flux criteria as described in \citet{coil15},
\citet{azadi17}, \citet{azadi18}, and \citet{leung19}.  Further
details on target selection, observations, spectroscopic data
reduction, slit loss corrections, and line flux measurements are
provided in several papers discussing results from the MOSDEF survey
(e.g., \citealt{kriek15, reddy15}).

\subsection{Stellar Population Modeling}
\label{sec:sedmodeling}

To aid in fitting the Balmer emission lines while accounting for
underlying Balmer absorption, we computed the stellar population
models that best fit the photometry of galaxies in our sample.  We fit
the \citet{bruzual03} (BC03) $Z=0.020$ ``solar'' metallicity stellar
population synthesis models to broadband photometry compiled in
\citet{skelton14}.  If applicable, the photometry was corrected for
the contribution from the strongest rest-frame optical emission lines
measured in the MOSFIRE spectra, including \oii, \hb, \oiii, \ha, and
\nii.  We assumed a constant star-formation history and ages that vary
from $50$\,Myr to the age of the Universe at the redshift of each
galaxy.  We considered stellar continuum reddening in the range
$0.0\le \ebmvstars \le 0.6$ for the \citet{reddy15} curve, which was
derived from MOSDEF galaxies targeted during the first two years of
the survey.  In Section~\ref{sec:discussion}, we also consider other
stellar attenuation curves that may be more applicable to galaxies in
the lower and upper halves of the stellar mass distribution of MOSDEF
galaxies.  The stellar mass, age, $\ebmvstars$ (i.e., reddening of the
stellar continuum), and SFR of the model that yields the lowest
$\chi^2$ relative to the photometry were taken to be the ``best-fit''
values, and the best-fit models themselves were used in fitting the
Balmer emission lines (Section~\ref{sec:ratiomeasurements}).

\subsection{Composite Spectra}
\label{sec:composites}

The weaker Balmer lines, e.g., $\hg$ and the higher-order transitions,
are typically not detected in the spectra of individual objects.  In
order to obtain robust constraints on the shape of the nebular
attenuation curve at wavelengths shorter than $\hg$, we constructed
composite spectra that allow us to detect these weaker Balmer lines.
The procedure for constructing these composite spectra proceeded as
follows.

\begin{figure}
\epsscale{1.20}
\plotone{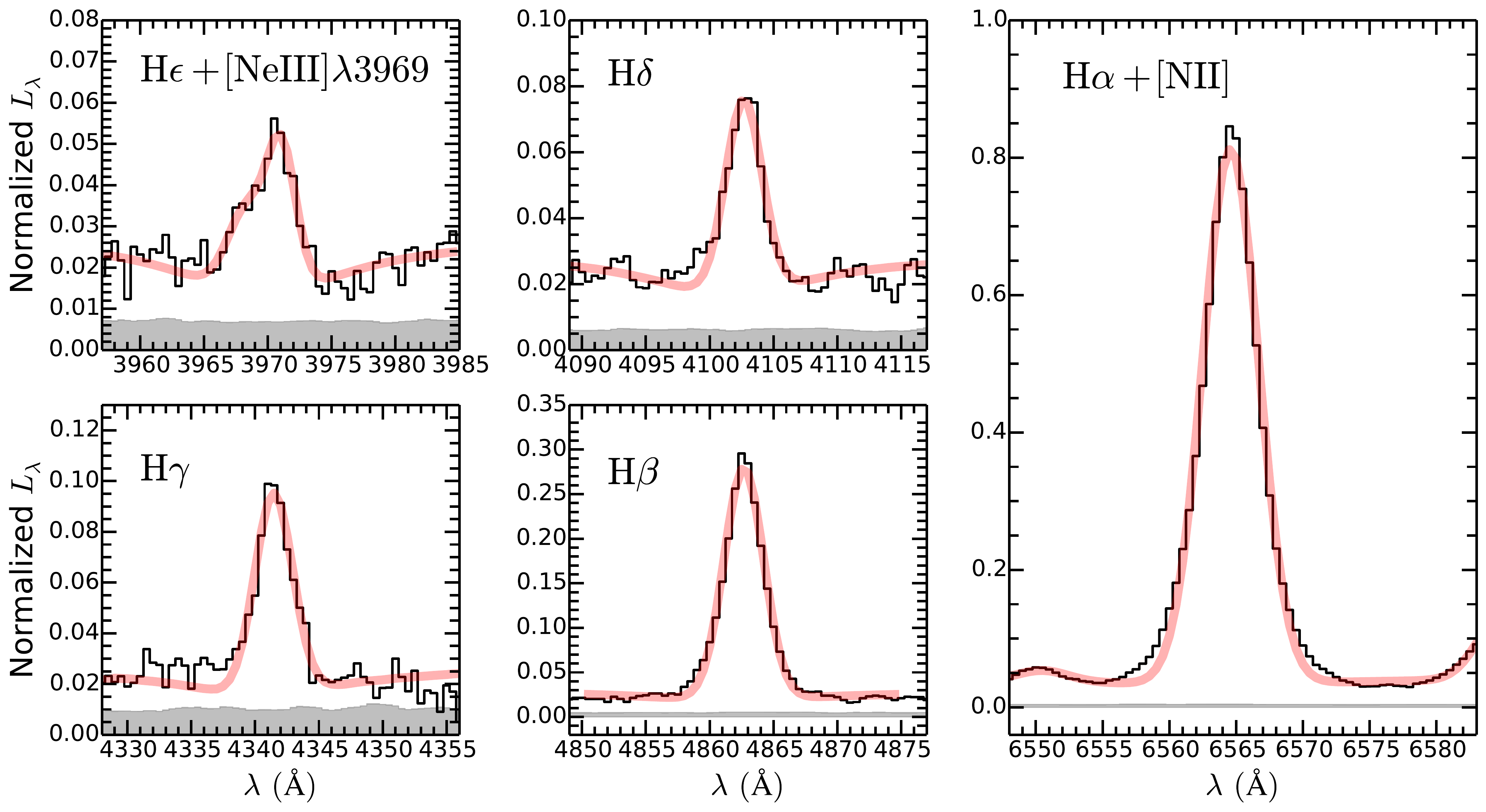}
\caption{Composite spectrum constructed for sample S4
  (Table~\ref{tab:samples}), shown in black, along with fits to each
  of the Balmer emission lines (red).  The error spectrum is indicated
  in grey.  For sample S4, only $\ha$, $\hb$, and $\he$ were used to
  compute the attenuation curve
  (Section~\ref{sec:sampleconstruction}).}
\label{fig:compfitexample}
\end{figure}

First, the science and error spectra were shifted to the rest-frame,
converted to luminosity density, and interpolated to a grid with a
wavelength spacing of $\delta\lambda_{\rm g} = 0.5$\,\AA.  Error
spectra were multiplied by $\sqrt{\delta\lambda_{\rm
    n}/\delta\lambda_{\rm g}}$ to account for resampling, where
$\delta\lambda_{\rm n}$ is the native wavelength spacing of the
spectra.  The science and error spectra were then normalized by the
$\ha$ luminosity measured from the science spectrum.  The composite
spectrum at each wavelength point was computed as an unweighted
average with 3$\sigma$ outlier rejection of the luminosity densities
of individual spectra at the same wavelength point.  The composite
error spectrum was calculated by adding in quadrature individual error
spectra and dividing by the total number of individual spectra
contributing to a given wavelength point.
Figure~\ref{fig:compfitexample} shows an example of one of the
composite spectra, along with the fits to the Balmer lines (see
Section~\ref{sec:ratiomeasurements}).  The average best-fit SED
corresponding to a composite spectrum was computed by simply averaging
the best-fit SEDs of individual galaxies contributing to that
composite spectrum.  This average best-fit SED aids in measuring the
Balmer lines, as discussed in the next section.

\subsection{Balmer Line Ratio Measurements from the Composite Spectra}
\label{sec:ratiomeasurements}

Average Balmer emission line ratios were computed from the composite
spectra as follows.  Each Balmer emission line was fit with one
Gaussian function (see below for exceptions) on top of a continuum
with a shape given by the average best-fit SED for that composite.  By
fixing the shape of the underlying continuum to that of the average
best-fit SED, we can account for underlying Balmer absorption and, in
the case of $\he$, additional absorption from the $\caii$ H line.  The
Balmer absorption line profiles in the average best-fit SEDs generally
agree well with the absorption line profiles observed in the
corresponding composite spectra (Figure~\ref{fig:compfitexample}).
The velocity widths of the Gaussian functions used to fit the Balmer
emission lines were fixed to lie within $20\%$ of the value obtained
for $\ha$.  We also allowed the centers of the Gaussian functions to
vary within $1.5$\,\AA\, of the rest-frame values listed in
Table~\ref{tab:lineproperties}.  The $\ha$ line was fit simultaneously
with the $\nii$ doublet using three Gaussian functions.\footnote{The
  composite $\ha$ line profile is slightly broadened relative to a
  Gaussian function.  Given the lack of a broadened component in the
  $\hb$, $\oii$, and $\oiii$ lines, we chose to fit a single Gaussian
  to the $\ha$ line.  Separately fitting the broadened component of
  this line results in a total $\ha$ luminosity that is $\simeq 1\%$
  larger than that obtained with a single Gaussian function.}

Additionally, $\he$ is blended with the
longer wavelength line of the $\neiii\lambda\lambda 3870,3969$
doublet.  The $\he+\neiii\lambda 3969$ blend is marginally resolved in
our spectra (Figure~\ref{fig:compfitexample}).  Thus, the $\he$ flux
was calculated by fitting simultaneously both the $\he$ and
$\neiii\lambda3969$ lines with two Gaussian functions, where the flux
of $\neiii\lambda3969$ was fixed to be $\approx 31\%$ of the flux of
$\neiii\lambda3870$, and the velocity width of $\neiii\lambda3969$ was
fixed to that of $\neiii\lambda3870$.

We considered only the first five low-order Balmer emission lines in
our analysis.  We did not use $\hz$ to calculate the dust curve as
this line is blended (and unresolved) with the $\hei\lambda 3889$
triplet line.  Moreover, the higher-order recombination lines are not
only intrinsically weaker than the lower-order lines, but, if dust is
present, they are progressively attenuated given their shorter
wavelengths.  $\hn$, $\htheta$, $\hiota$, and all other higher-order
Balmer emission lines, as well as the higher-order Paschen lines that
have coverage in the MOSFIRE spectra (e.g., P$\delta$ and higher for
galaxies at $z\sim 1.4$, P$\iota$ and higher at $z\sim 2.0$, and so
on), were not considered in our analysis as none of these lines are
detected with $S/N\ge 3$ in any of the composite spectra considered
here.

Each Balmer emission line was fit by considering only those wavelength
points lying in the windows specified in
Table~\ref{tab:lineproperties}, and allowing the continuum
normalization to vary.  Because the individual galaxy spectra were
normalized by the $\ha$ fluxes before combining them into composites
(Section~\ref{sec:composites}), the line measurements obtained from
these composites represent the average ratios of the line luminosities
to that of $\ha$.  All of the fitting described above, along with the
line flux error calculations, were implemented using IDL's MPFIT
package \citep{markwardt09}.

\section{\bf Sample Construction for Nebular Attenuation Curve Analysis}
\label{sec:sampleconstruction}

The most robust and useful constraints on the shape of the nebular
dust attenuation curve can be obtained from large samples of galaxies
with coverage of as many $\hi$ (or other) recombination emission lines
as possible.  Large samples allow us to measure the average
emission-line fluxes with greater precision and detect weaker
recombination lines, while coverage of many lines allows us to
determine the shape of the dust curve over a broad range of
wavelengths.

\begin{figure}
\epsscale{1.0}
\plotone{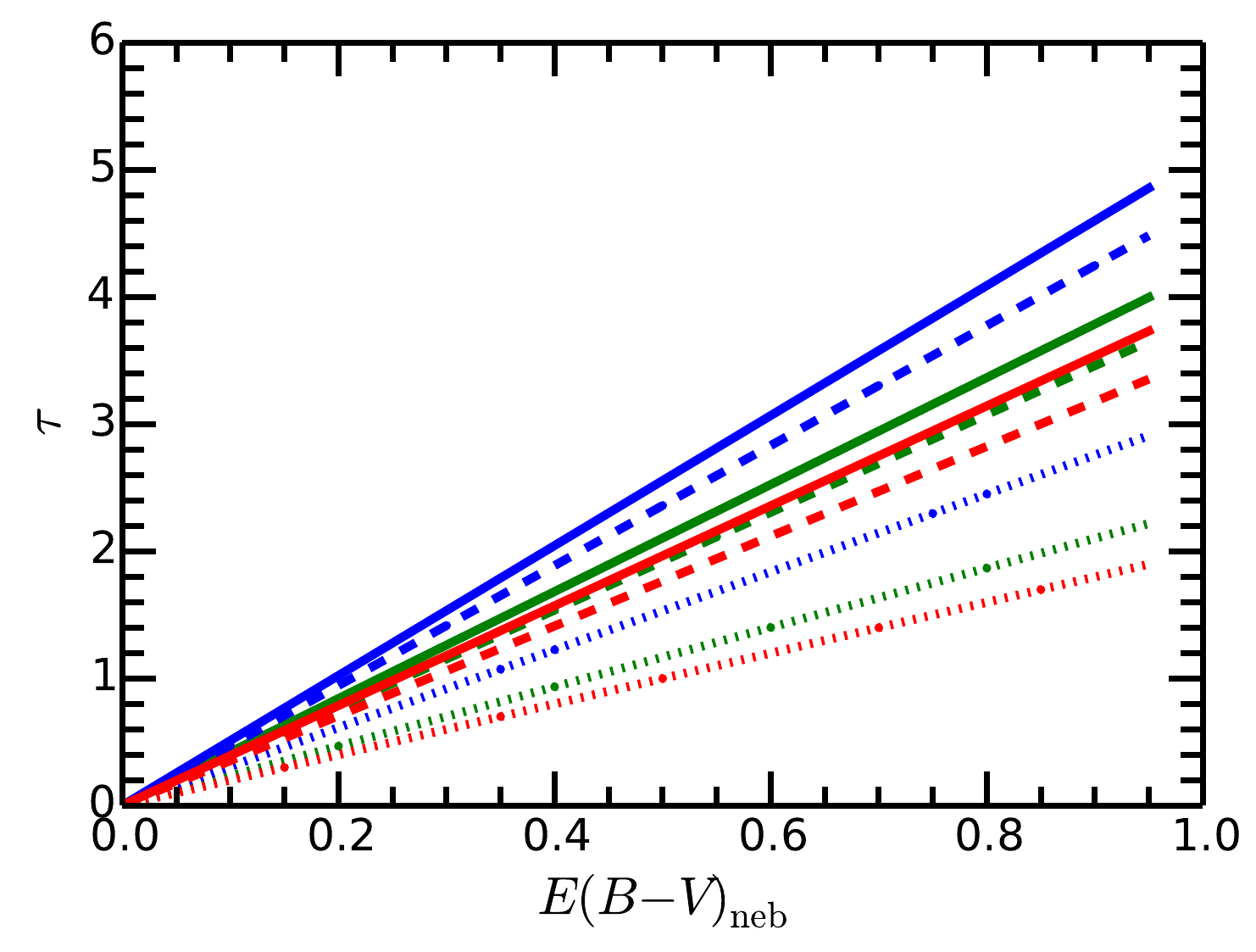}
\caption{Optical depth ($\tau$) of $\he$ (solid lines), $\hg$
  (dashed lines), and $\ha$ (dotted lines) versus $\ebmvgas$ for the
  \citet{calzetti00} attenuation curve (blue), SMC extinction curve
  (red), and the MW curve (green).}
\label{fig:tau}
\end{figure}

The method of using multiple recombination line ratios to constrain
the shape of the nebular dust attenuation curve relies on the
assumption that the optical depths ($\tau$) of all of the line
transitions are less than a few.  In particular, for very dusty
galaxies, the variation in optical depth with wavelength could result
in a situation where the high-order Balmer recombination lines are
dominated by emission from the relatively unobscured regions of a
galaxy, while the low-order lines are dominated by emission from the
dustier and potentially physically distinct regions of a galaxy.  In
this scenario, one would deduce less attenuation at bluer wavelengths.
The average $\ebmvgas$ of galaxies in our sample, derived assuming the
Galactic (Milky Way, or MW) extinction curve (\citealt{cardelli89}),
is $\langle \ebmvgas\rangle \simeq 0.35$, with $\ga 90\%$ of the
galaxies having $\ebmvgas < 0.7$.  Based on this $\ebmvgas$
distribution, we expect $\tau(\he) \la 2$ for typical galaxies in our
sample, with the vast majority having $\tau(\he) \la 4$
(Figure~\ref{fig:tau}), suggesting that the wavelength variation in
optical depth is unlikely to affect our determination of the nebular
dust attenuation curve.\footnote{The attenuation curve determined from
  only those galaxies that have $\ebmvgas<1$, as calculated with the
  MW curve, is identical within the errors to the one found using the
  entire sample.}

Nevertheless, when using ratios of these recombination lines to
compute the shape of the dust curve, we must take care that the
distribution of nebular reddening of individual galaxies contributing
to each line is identical.  Simply put, we must ensure that exactly
the same set of galaxies contributes to each of the lines.  As the
galaxies in our sample span a range of redshifts and may have been
observed in different sets of filters, not all galaxies that have
coverage of one emission line will have coverage in others.

To ensure that the same set of galaxies contributes to each line used
to compute the nebular attenuation curve, we subdivided our sample
into sets of galaxies that have various combinations of covered
emission lines in common.  We identified 9 samples of galaxies with
unique combinations of Balmer emission lines with spectral coverage.
The characteristics of these samples are listed in
Table~\ref{tab:samples}, including the number of galaxies belonging to
each sample, the redshift range and mean redshift of galaxies in each
sample, and the fluxes of the covered emission lines that all the
galaxies in each sample have in common.  Because the calculated $\he$
flux depends on that measured for $\neiii\lambda 3870$
(Section~\ref{sec:ratiomeasurements}), a galaxy is considered to have
coverage of $\he$ only if it also has coverage of $\neiii\lambda3870$.

\begin{deluxetable*}{lrlcccccl}
\tabletypesize{\footnotesize}
\tablewidth{0pc}
\tablecaption{Samples, Balmer Emission-Line Coverage, and Line Flux Ratios}\tablenotemark{a}
\tablehead{
\colhead{Sample} &
\colhead{$N$\tablenotemark{b}} &
\colhead{$z$-Range ($\langle z\rangle$)\tablenotemark{c}} &
\colhead{$\ha$} &
\colhead{$\hb$} &
\colhead{$\hg$} &
\colhead{$\hd$} &
\colhead{$\he$\tablenotemark{d}} &
\colhead{Subset of\tablenotemark{e}}}
\startdata

 S1 & 240 & $1.2467-2.6403$ ($1.9424$) & {\bf $ 1.000 \pm 0.004$} & {\bf $ 0.234 \pm 0.004$} & {\bf $ 0.085 \pm 0.004$} & --- & --- & --- \\
 S2 & 130 & $1.3631-2.6196$ ($1.5875$) & {\bf $ 1.000 \pm 0.004$} & {\bf $ 0.224 \pm 0.004$} & {\bf $ 0.085 \pm 0.004$} & {\bf $ 0.038 \pm 0.004$} & --- & S1, S7 \\
 S3 &  72 & $1.5031-2.1230$ ($1.6719$) & {\bf $ 1.000 \pm 0.005$} & {\bf $ 0.225 \pm 0.006$} & {\bf $ 0.076 \pm 0.006$} & {\bf $ 0.033 \pm 0.005$} & {\bf $ 0.020 \pm 0.007$} & S1-S2, S4-S9 \\
 S4 & 355 & $1.5031-2.4225$ ($2.1041$) & {\bf $ 1.000 \pm 0.003$} & {\bf $ 0.247 \pm 0.003$} & --- & --- & {\bf $ 0.024 \pm 0.004$} & --- \\
 S5 &  80 & $1.5031-2.4225$ ($1.7433$) & {\bf $ 1.000 \pm 0.005$} & {\bf $ 0.227 \pm 0.006$} & {\bf $ 0.077 \pm 0.006$} & --- & {\bf $ 0.021 \pm 0.007$} & S1, S4 \\
 S6 & 278 & $1.5031-2.3104$ ($2.0384$) & {\bf $ 1.000 \pm 0.003$} & {\bf $ 0.247 \pm 0.003$} & --- & {\bf $ 0.044 \pm 0.003$} & {\bf $ 0.023 \pm 0.004$} & S4, S9 \\
 S7 & 141 & $1.3549-2.6196$ ($1.6206$) & {\bf $ 1.000 \pm 0.004$} & --- & {\bf $ 0.085 \pm 0.004$} & {\bf $ 0.038 \pm 0.004$} & --- & --- \\
 S8 &  82 & $1.5031-2.1244$ ($1.7213$) & {\bf $ 1.000 \pm 0.005$} & --- & {\bf $ 0.077 \pm 0.006$} & {\bf $ 0.033 \pm 0.005$} & {\bf $ 0.019 \pm 0.007$} & S7, S9 \\
 S9 & 289 & $1.5031-2.3104$ ($2.0401$) & {\bf $ 1.000 \pm 0.003$} & --- & --- & {\bf $ 0.044 \pm 0.003$} & {\bf $ 0.022 \pm 0.004$} & --- 
\enddata
\tablenotetext{a}{Sample characteristics and line fluxes relative to $\ha$.  
For each sample, flux measurements are included only for those
lines that have $S/N \ge 3$ in the composite spectrum and have coverage for all galaxies in that sample.  Only lines that satisfy these requirements are 
used to compute the attenuation curve for each sample.}
\tablenotetext{b}{Number of galaxies in the sample.}
\tablenotetext{c}{Redshift range and the mean redshift (in parentheses) of galaxies
in the sample.  Because of the multi-modal redshift distributions of some of the samples,
there may be few galaxies that lie at the mean redshift of galaxies in the sample.}
\tablenotetext{d}{A galaxy is considered to have coverage of $\he$ only if
its spectrum also includes $\neiii\lambda 3870$, as the latter was used to
estimate the contribution of $\neiii\lambda 3969$ to the $\he$+$\neiii\lambda 3969$
blend (Section~\ref{sec:ratiomeasurements}).  The $\he/\ha$ line flux ratios reported in this column are corrected for the contribution
of $\neiii\lambda 3969$.}
\tablenotetext{e}{Indicates the samples of which this sample is a subset.}
\label{tab:samples}
\end{deluxetable*}

In constructing these samples, we required that the longest wavelength
line with coverage, namely $\ha$, be detected with $S/N\ge 3$, so that
the individual galaxy spectra may be normalized by the $\ha$ flux when
calculating the composite spectrum
(Section~\ref{sec:composites}).\footnote{The MOSDEF sample includes a
  number of galaxies at $z\ga 3$ that do not have coverage of $\ha$,
  but that do have high $S/N$ detections of $\hb$.  While the
  attenuation curve can, in principle, be derived for these galaxies,
  we chose not to include them in our analysis as the lack of $\ha$
  coverage results in larger uncertainties in $\ebmvgas$ and,
  consequently, the shape of the attenuation curve.}  There is no
requirement for coverage placed on lines that have blank entries in
Table~\ref{tab:samples}.  As such, the different samples listed in
Table~\ref{tab:samples} have, in general, many galaxies in common.
For example, sample S3 consists of all those galaxies in sample S2
that have coverage of $\he$, in addition to all the lines required to
be covered in S2.  {The last column of Table~\ref{tab:samples}
  indicates for each sample all of the other samples of which it is a
  subset of.}  In total, we used 532 galaxies from the MOSDEF sample
to constrain the shape of the nebular attenuation curve.

Composite spectra were constructed for each of the 9 samples listed in
Table~\ref{tab:samples}.  For each composite, we measured the covered
Balmer emission lines associated with the corresponding sample.  These
measurements were used to compute the shape of the nebular attenuation
curve for each sample, as discussed in the next section.

\section{\bf CALCULATION OF THE SHAPE OF THE NEBULAR ATTENUATION CURVE}
\label{sec:calculations}

The relationship between the observed (or attenuated) and intrinsic
fluxes of a line centered at wavelength $\lambda$, denoted by
$f(\lambda)$ and $f_{\rm 0}(\lambda)$, respectively, can be expressed as
follows:
\begin{eqnarray}
f(\lambda) = f_{\rm 0}(\lambda) \times 10^{-0.4A(\lambda)},
\label{eq:observedintrinsic}
\end{eqnarray}
where $A(\lambda)$ is the attenuation in magnitudes at wavelength $\lambda$.
For any two lines centered at wavelengths $\lambda_{\rm 1}$ and $\lambda_{\rm 2}$,
we can then write
\begin{eqnarray}
\frac{f(\lambda_{\rm 1})}{f(\lambda_{\rm 2})} = \frac{f_{\rm 0}(\lambda_{\rm 1})}{f_{\rm 0}(\lambda_{\rm 2})} \times 10^{-0.4[A(\lambda_{\rm 1}) - A(\lambda_{\rm 2})]},
\end{eqnarray}
or 
\begin{eqnarray}
A(\lambda_{\rm 2}) & = & 2.5\left[\log_{\rm 10}\left(\frac{f(\lambda_{\rm 1})}{f(\lambda_{\rm 2})}\right) - \log_{\rm 10}\left(\frac{f_{\rm 0}(\lambda_{\rm 1})}{f_{\rm 0}(\lambda_{\rm 2})}\right)\right] \nonumber \\
& & + A(\lambda_{\rm 1}).
\label{eq:alam2}
\end{eqnarray}
In our analysis, $\lambda_{\rm 1}$ denotes the
wavelength of the reddest line with coverage, namely $\ha$
(Table~\ref{tab:samples}).  We define a new quantity,
\begin{eqnarray}
A'(\lambda_{\rm 2}) & \equiv & A(\lambda_{\rm 2}) + [1-A(\lambda_{\rm 1})] \nonumber \\
& = & 2.5\left[\log_{\rm 10}\left(\frac{f(\lambda_{\rm 1})}{f(\lambda_{\rm 2})}\right) - \log_{\rm 10}\left(\frac{f_{\rm 0}(\lambda_{\rm 1})}{f_{\rm 0}(\lambda_{\rm 2})}\right)\right] + 1,
\label{eq:aprime}
\end{eqnarray}
which is equivalent to $A(\lambda_{\rm 2})$ for $A(\lambda_{\rm
  1})=1$, and which depends only on measured line flux ratios.  With
these definitions, there is an offset between $A'(\lambda_{\rm 2})$
and $A(\lambda_{\rm 2})$, $1-A(\lambda_{\rm 1})$, that is constant and
independent of $\lambda_{\rm 2}$.  The intrinsic Balmer emission line
ratios (Table~\ref{tab:lineproperties}), along with the observed
ratios measured from the composites constructed for the samples listed
in Table~\ref{tab:samples}, were used with Equation~\ref{eq:aprime} to
compute $A'(\lambda_{\rm 2})$.  We then fit $A'(\lambda_{\rm 2})$
versus $\lambda_{\rm 2}$ using linear and quadratic polynomials of the
form $A'(\lambda) = a_{\rm 0} + a_{\rm 1}/\lambda$ (linear in
$1/\lambda$) and $A'(\lambda) = a_{\rm 0} + a_{\rm 1}/\lambda + a_{\rm
  2}/\lambda^2$ (quadratic in $1/\lambda$), respectively, where
$\lambda$ is the wavelength in $\mu$m.  While the
quadratic-in-$1/\lambda$ form results in reduced $\chi^2$ that are
typically an order of magnitude smaller than those obtained with the
linear-in-$1/\lambda$ form, we present results using both forms to
demonstrate the degree to which the functional fit affects the derived
attenuation curve.  Our measurements lack sufficient precision and
wavelength sampling to warrant more complicated functional forms for
the wavelength dependence of the attenuation curve.

The attenuation curve is defined as
\begin{eqnarray}
k(\lambda) \equiv \frac{A(\lambda)}{E(B-V)},
\label{eq:kae}
\end{eqnarray}
where $\ebmv = A(B)-A(V)$ is the reddening in magnitudes.  As we are
concerned with the {\em nebular} attenuation curve, $\ebmv$ is the
reddening appropriate to the ionized gas, $\ebmvgas$.  Taking the
effective wavelengths of the $B$ and $V$ bands to be $4400$\,\AA\, and
$5500$\,\AA, respectively, we can then define an attenuation curve
that is related to $A'(\lambda_{\rm 2})$:
\begin{eqnarray}
k^\prime(\lambda) = \frac{A'(\lambda)}{A'(4400\,{\rm \AA}) - A'(5500\,{\rm \AA})},
\label{eq:kprime}
\end{eqnarray}
where we have set $\lambda=\lambda_{\rm 2}$ for simplicity.  Given the
previous definitions, the offset between $k^\prime(\lambda)$ and
$k(\lambda)$ is $[1-A(\lambda_{\rm 1})]/\ebmvgas$ and is independent
of $\lambda$.  Accordingly, determining $k^\prime(\lambda)$ is
effectively equivalent to determining $k(\lambda)$ up to a
normalization constant (Section~\ref{sec:normofcurve}).

In summary, we used the observed and intrinsic Balmer emission line
ratios to compute $A'(\lambda_{\rm 2})$ using
Equation~\ref{eq:aprime}, fit linear and quadratic polynomials to
$A'(\lambda_{\rm 2})$, used these polynomial fits to determine the
values of $A'(4400\,{\rm \AA})$ and $A'(5500\,{\rm \AA})$, and then
used Equation~\ref{eq:kprime} to calculate $k^\prime(\lambda)$.  As
noted above, $k^\prime(\lambda)$ is equivalent to $k(\lambda)$ apart
from a normalization constant.  Measurement uncertainties were
propagated throughout these calculations, such that the final
uncertainty in a given $k^\prime(\lambda)$ point includes uncertainty
in line flux measurements, slit loss corrections for those line ratios
where the two lines were observed in different bands\footnote{The
  random uncertainty in a line flux ratio due to slit loss
  corrections, where the two lines were observed in different bands,
  is $\sim 18\%$ (see Section~\ref{sec:slitlosscorrections}).  The
  total uncertainty due to slit loss corrections when measuring lines
  in the composite spectra is set equal to $18\% \sqrt{N_{\rm
      1}}/N_{\rm 2}$, where $N_{\rm 1}$ is the number of objects
  contributing to the composite for which the two lines of the line
  ratio are covered in different filters and $N_{\rm 2}$ is the total
  number of objects contributing to the composite.}, and $\ebmvgas$.
All of the $k^\prime(\lambda)$ points determined for individual
samples were then fit together, weighted by their inverse variances,
using the polynomial forms discussed above to produce a final
attenuation curve.  In practice, the final attenuation curve was
determined using $k^\prime(\lambda)$ points from only those lines that
had $S/N \ge 3$.

\begin{figure*}
\epsscale{1.15}
\plotone{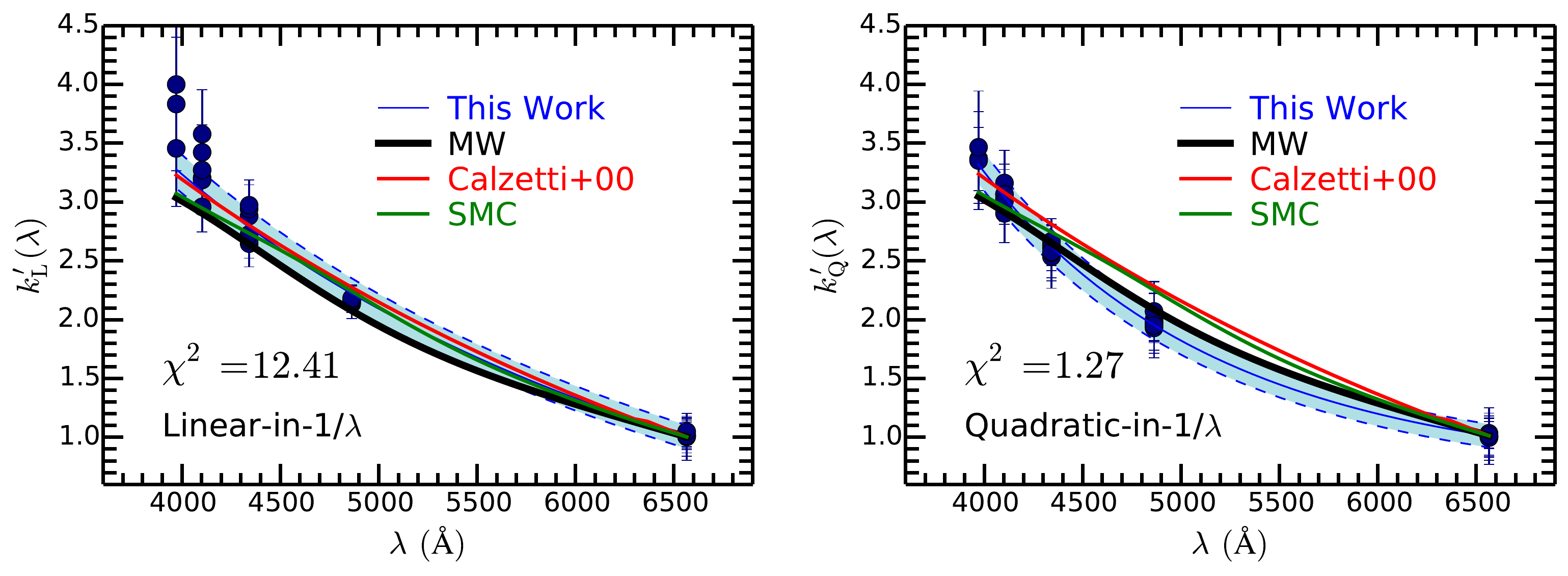}
\caption{$k^\prime(\lambda)$ versus $\lambda$ for the
  linear-in-$1/\lambda$ ({\em left}) and quadratic-in-$1/\lambda$
  polynomial forms ({\em right}).  Measurements are shown by the blue
  circles---these values differ between the two polynomial forms since
  $k^\prime(\lambda)$ depends on $\ebmvgas$ and the latter depends on
  the functional form used to fit $A'(\lambda)$.  Error bars for each
  $k^\prime(\lambda)$ point are also shown.  The best-fit polynomial
  and $95\%$ confidence intervals are denoted by the solid and dashed
  blue lines, respectively.  The reduced $\chi^2$ for the fits are indicated
  in each panel.  For comparison, the MW extinction curve, shifted so
  that its value at the wavelength of $\ha$ is equal to that of the
  curve derived here, is shown by the thick black line.  Similarly,
  the shifted SMC and \citet{calzetti00} curves are shown by the solid
  green and red lines, respectively.  }
\label{fig:kave}
\end{figure*}

Because there are many galaxies in common between the samples listed
in Table~\ref{tab:samples} (Section~\ref{sec:sampleconstruction}), the
$k^\prime(\lambda)$ points computed from these samples are not
completely independent of each other.  Thus, the formal uncertainty in
the fit to $k^\prime(\lambda)$ will underestimate the true measurement
uncertainty.  To determine the error in the fit to
$k^\prime(\lambda)$, the samples were restricted so that all galaxies
in each sample have exactly the same set of covered emission lines
(i.e., any galaxies that have coverage of lines with blank entries in
Table~\ref{tab:samples} are excluded from the samples).  With this
requirement, the individual samples do not have any galaxies in common
and will contain fewer galaxies than indicated in
Table~\ref{tab:samples}---all but five samples (S1, S2, S3, S4, and
S6) contain fewer than 10 galaxies.
The $k^\prime(\lambda)$ points from these five samples were fit using
linear and quadratic polynomials as described above, and we adopted the
formal uncertainty on these fits as representative of the actual
measurement uncertainty in the mean nebular attenuation curve.
The functional forms
of $k^\prime(\lambda)$ versus $1/\lambda$ are
\begin{eqnarray}
k^\prime_{\rm L}(\lambda) & = & -2.479 + \frac{2.286}{\lambda}
\label{eq:kcconstlin}
\end{eqnarray}
for the linear-in-$1/\lambda$ fit and 
\begin{eqnarray}
k^\prime_{\rm Q}(\lambda) & = & 2.074 - \frac{2.519}{\lambda} + \frac{1.196}{\lambda^2}
\label{eq:kcconst}
\end{eqnarray}
for the quadratic-in-$1/\lambda$ fit 
for $0.38\la \lambda \la 0.66$\,$\mu$m. 
The curves have been shifted in normalization so that $k^\prime_{\rm
  L,Q}(\ha) = 1$.  The subscripts ``L'' and ``Q'' refer to the
linear-in-$1/\lambda$ and quadratic-in-$1/\lambda$ forms,
respectively, and are used to distinguish the curves found here from
other common extinction/attenuation curves.

\section{\bf RESULTS}
\label{sec:results}

The attenuation curve derived here is shown in Figure~\ref{fig:kave},
along with the MW, \citet{calzetti00}, and SMC curves, all shifted in
normalization so that $k^\prime(\ha)_{\rm MW,Calz,SMC} = k^\prime_{\rm
  L,Q}(\ha)$.  The MW (Galactic extinction) curve is most commonly
used to derive the nebular reddening and dust corrections to nebular
lines, while the \citet{calzetti00} and SMC curves are typically
assumed for the reddening of the stellar continuum in high-redshift
galaxies.  In what follows, we compare the curve found here with
the Galactic extinction curve and several other commonly used
extinction/attenuation curves.

\subsection{Shape of the Average Nebular Attenuation Curve}
\label{sec:shapeofcurve}

Our results imply that the shape of the nebular attenuation curve is
similar to that of the MW within the uncertainties, irrespective of
the functional (linear or quadratic) form adopted for the curve, as
shown in Figure~\ref{fig:kave}.  This figure also highlights the
systematic uncertainty stemming from the adopted functional form for
$k^\prime(\lambda)$.  Specifically, $k^\prime(\lambda)$ depends on
$\ebmvgas$, and $\ebmvgas$ is derived from a polynomial fit to
$A'(\lambda)$ (Section~\ref{sec:calculations}).  The
linear-in-$1/\lambda$ fit to $A'(\lambda)$ results in $\ebmvgas$ that
is $\approx 17\%$ smaller than that derived from the
quadratic-in-$1/\lambda$ fit.  As a result, the $k^\prime(\lambda)$
points for the linear-in-$1/\lambda$ function are the same percentage
larger than those derived for the quadratic-in-$1/\lambda$ function.

While this systematic uncertainty does not change our conclusion that
the curve found here is similar to the Galactic extinction curve, it
does affect the comparison of the curve to those that are commonly
adopted to dust-correct the stellar continuum emission in galaxies.
In particular, the simple linear-in-$1/\lambda$ form implies a nebular
attenuation curve that is indistinguishable in shape from other common
curves at rest-frame optical wavelengths.  On the other hand, adopting
the quadratic-in-$1/\lambda$ form yields a nebular attenuation curve
that has more curvature than other common curves at rest-frame optical
wavelengths.

There are a couple of reasons why the quadratic-in-$1/\lambda$ form
may be preferred.  First, this functional form yields a significantly
lower reduced $\chi^2$ than the linear-in-$1/\lambda$ form.  The latter
predicts lower values of $k^\prime(\lambda)$ at $\lambda \la
4500$\,\AA, relative to $\ha$, than what the shorter wavelength Balmer
lines suggest.  Second, most of the common extinction/attenuation
curves found in the literature (e.g., \citealt{cardelli89, gordon03,
  calzetti00} require a higher-order polynomial to fully capture their
shape at rest-frame optical wavelengths, in contrast to the
longer-wavelength ($\lambda \ga 7000$\,\AA) behavior that is typically
parameterized by an inverse power-law in $\lambda$.  Precise
measurements of even higher-order Balmer lines and lines of the
Paschen series, as well as the underlying stellar absorption, will
clarify the functional form of the nebular attenuation curve over a
broader range of wavelengths than is currently accessible.  Finally,
whether the aforementioned similarities and differences between the
various curves extend to the reddening at UV wavelengths where the
variations between extinction/attenuation curves are more pronounced
remains unknown.

\subsection{Normalization of the Average Nebular Attenuation Curve}
\label{sec:normofcurve}

The normalization of the total nebular dust attenuation curve,
$k(\lambda)$, may be found by extrapolating $k^\prime(\lambda)$ to
some sufficiently long wavelength and setting the curve to be zero at
this point (e.g., \citealt{reddy15}).  The MW, SMC, and LMC curves all
become very close to zero at $\lambda \ga 2.8$\,$\mu$m.  Extrapolating
Equation~\ref{eq:kcconstlin} to $2.8$\,$\mu$m and forcing the value to
be zero at this point implies the following total attenuation curve for
the linear-in-$1/\lambda$ fit: 
\begin{eqnarray}
k_{\rm L}(\lambda) & = & -0.816 + \frac{2.286}{\lambda}, \nonumber \\
& & 0.40 \le \lambda \le 0.66\, \mu{\rm m}.
\label{eq:kclin}
\end{eqnarray}
A similar normalization of the quadratic-in-$1/\lambda$ fit requires one to
force
$k^\prime_{\rm Q}(\lambda)$ (Equation~\ref{eq:kcconst}) to conform to
the commonly-adopted behavior where $k(\lambda)\propto 1/\lambda$ as
$\lambda \rightarrow \infty$.  Thus, we shifted $k^\prime_{\rm Q}(\lambda)$ to have
the same value as $k_{\rm L}(\lambda)$ at 0.66\,$\mu$m to ensure a continuous function,
thus obtaining the following total
attenuation curve for the quadratic-in-$1/\lambda$ fit:
\begin{eqnarray}
k_{\rm Q}(\lambda) & = & 3.719 - \frac{2.519}{\lambda} + \frac{1.196}{\lambda^2}, \nonumber \\
& & 0.40 \le \lambda \le 0.66\, \mu{\rm m}; \nonumber \\
& = & -0.816 + \frac{2.286}{\lambda}, \nonumber \\
& & \lambda > 0.66\, \mu{\rm m}.
\label{eq:kcquad}
\end{eqnarray}
The extrapolation of the total attenuation curve redward of $\lambda =
0.66$\,$\mu{\rm m}$ should be used with caution given that $k_{\rm
  L}(\lambda)$ is constrained only using points blueward of this
limit.  The ratio of the total-to-selective absorption at {\em V}-band
is $R_V=3.34$ and $3.09$ for the linear and quadratic forms of the
total attenuation curve, respectively. The difference between these
values ($\delta R_V\simeq 0.2$) gives an estimate of the systematic
uncertainty in $R_V$ that stems from assuming different functional
forms of $k(\lambda)$ versus $\lambda$ at long wavelengths.  There is
additional uncertainty associated with the specific wavelength at
which the attenuation curve is forced to zero.  For example, the
difference in $R_V$ obtained when assuming a ``zero wavelength'' of
$3.0$\,$\mu$m rather than $2.8$\,$\mu$m is $\delta R_V \simeq 0.05$.
The values of $R_V$ obtained here are entirely consistent with that of
the Galactic extinction curve ($R_V = 3.1$; \citealt{cardelli89})
given the aforementioned systematic errors.

\subsection{Balmer Emission Line Ratios of Individual Galaxies}

\begin{figure}
\epsscale{1.0}
\plotone{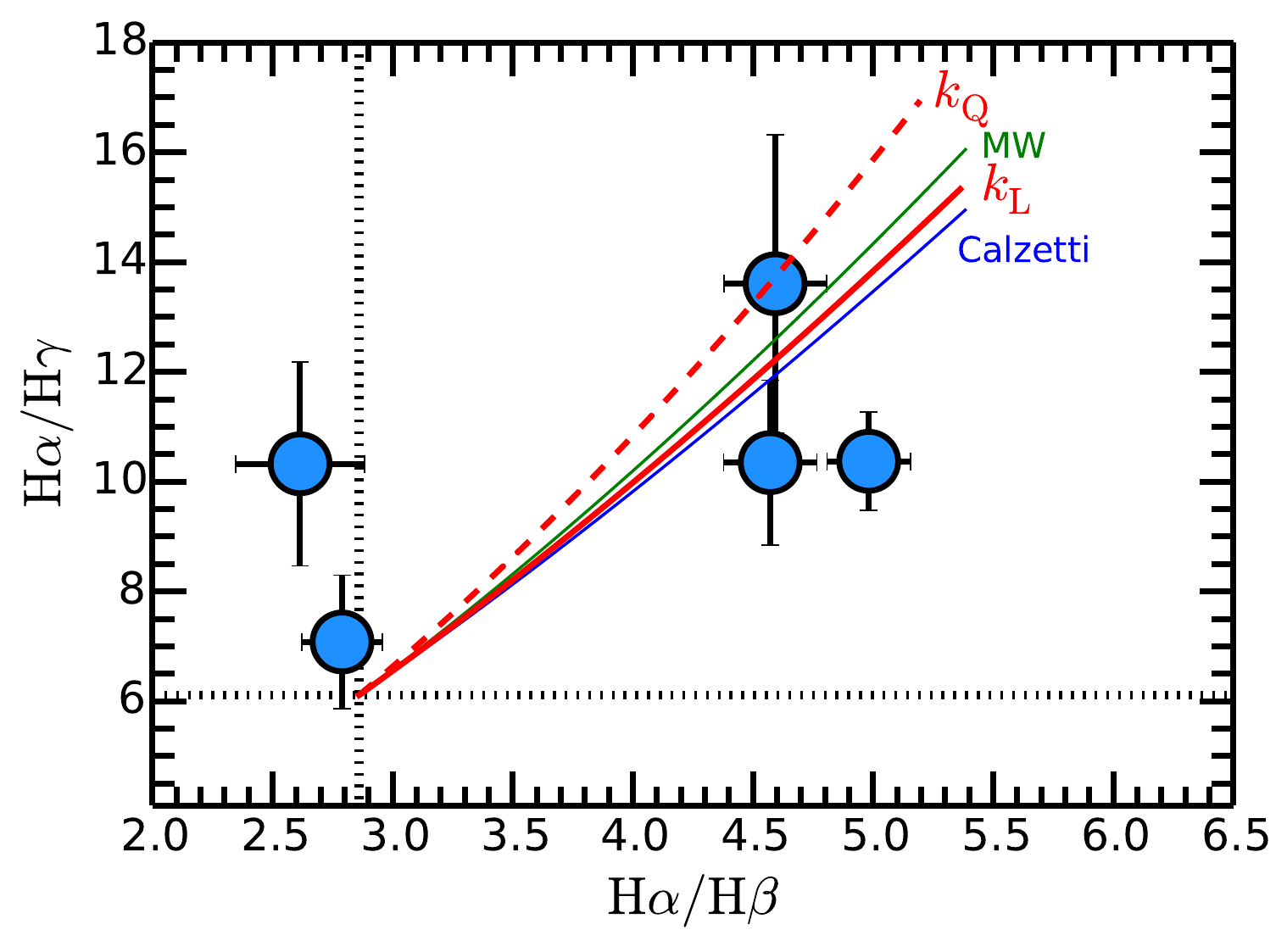}
\caption{Ratios of $\ha/\hg$ versus $\ha/\hb$ for five galaxies where
  $\ha$, $\hb$, and $\hg$ are detected with $S/N \ge 5$ (points).  The
  relationships between these ratios for different attenuation curves
  are indicated by the curves, including the linear and quadratic
  forms given by Equations~\ref{eq:kclin} and \ref{eq:kcquad}, where
  reddening increases towards the upper right-hand side of the figure.
  The SMC curve lies very close to that of the \citet{calzetti00}
  curve, and the \citet{reddy15} curve lies very close to $k_{\rm
    L}(\lambda)$, on this figure.  The dotted lines indicate the
  intrinsic line ratios.  The two objects whose ratios lie below the
  intrinsic values are consistent within $3\sigma$ of having very
  little reddening.}
\label{fig:indbalm}
\end{figure}

There are a handful of galaxies where the higher-order Balmer emission
lines (e.g., $\hg$ and $\hd$) are detected in individual spectra.  The
joint uncertainties in the resulting multiple Balmer emission line
ratios (i.e., $\ha/\hb$, $\ha/\hg$, etc.) are such that we cannot rule
out any of the nebular dust attenuation curves described above.
Figure~\ref{fig:indbalm} shows the $\ha/\hb$ and $\ha/\hg$ ratios of
the five galaxies in our sample where $\ha$, $\hb$, and $\hg$ are
detected with $S/N\ge 5$, relative to how these ratios depend on each
other for different attenuation/extinction curves.  These measurements
were made by simultaneously fitting the stellar continuum from the
best-fit SED model with the spectra of the individual line detections,
thus accounting for underlying Balmer absorption.  The measurements on
average are consistent within the errors with all of the
aforementioned extinction/attenuation curves.

\section{\bf DISCUSSION}
\label{sec:discussion}

The nebular attenuation curve has an important bearing on the
derivation of a number of fundamental physical properties of galaxies
and the ISM contained within them.  Our results imply a nebular
attenuation curve that is similar to the Galactic extinction curve
within the random and systematic uncertainties discussed above.  Here,
we consider the nebular reddening of galaxies in our sample, how the
nebular reddening compares to the reddening of the stellar continuum,
and how differences between the two values of reddening may be
interpreted in the context of recent advancements in stellar
population synthesis modeling.

\subsection{Reddening Comparisons}
\label{sec:reddening}

\begin{figure}
\epsscale{1.0}
\plotone{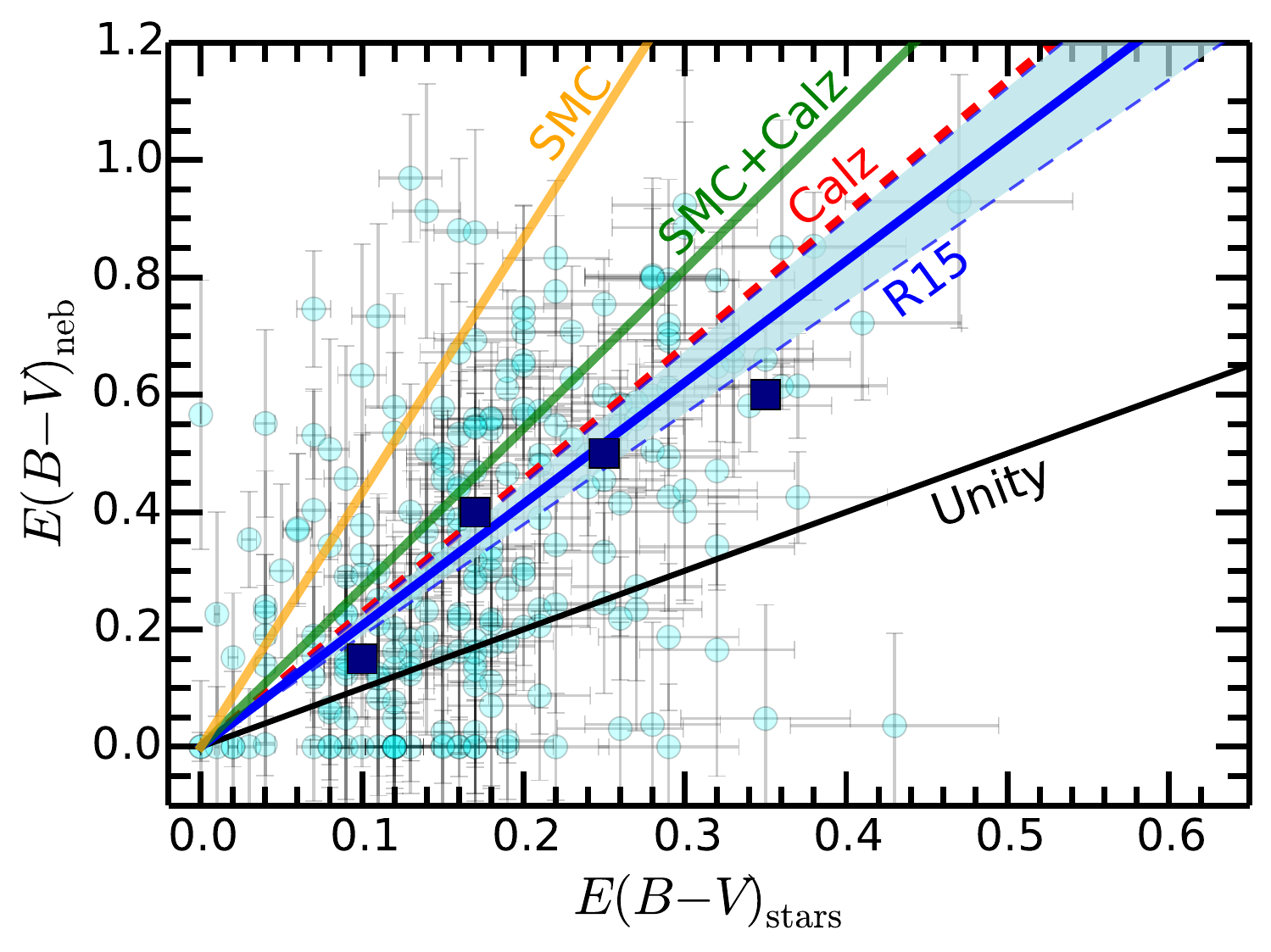}
\caption{Nebular reddening, derived assuming the $k_{\rm Q}(\lambda)$
  parameterization of the nebular attenuation curve, versus the
  reddening of the stellar continuum for individual galaxies (circles)
  with the fiducial modeling described in
  Section~\ref{sec:sedmodeling}, i.e., assuming the
    \citet{reddy15} (R15) attenuation curve.  The large squares
  indicate the average $\ebmvgas$ determined from composite spectra
  that were constructed in bins of $\ebmvstars$ and include objects
  where $\hb$ is not significantly detected.  The one-to-one (unity)
  relation is shown by the black line.  The best-fit linear relation
  to individual points and $95\%$ confidence intervals are indicated
  by the solid blue line and shaded regions, respectively.
  Additionally, we show the relations obtained if we assume the
  \citet{calzetti00} relation (dashed red line), an SMC extinction
  curve (solid orange line), and combined modeling that assumes the
  SMC and \citet{calzetti00} curves for low- and high-mass galaxies,
  respectively (see text).  In all cases, we find that $\ebmvgas$ is
  larger than $\ebmvstars$ on average.}
\label{fig:ebmvcompare}
\end{figure}

The nebular reddening, $\ebmvgas$, can be derived from the Balmer
decrement, $\ha/\hb$, as follows:
\begin{equation}
\ebmvgas = \frac{2.5}{k(\hb)-k(\ha)}\log_{10}\left[\frac{\ha/\hb}{2.86}\right].
\end{equation}
The similarity between the nebular attenuation curve found here and
the Galactic extinction curve imply that the corresponding $\ebmvgas$
are also similar.  Motivated by the expected correlation between the
reddening of the ionized gas and stellar continuum in galaxies (e.g.,
\citealt{fanelli88, calzetti97, calzetti00, kreckel13}), we recomputed
the relationship between $\ebmvgas$ and $\ebmvstars$, where the former
was computed assuming $k_{\rm Q}(\lambda)$ and the latter was
determined from SED-fitting for our fiducial assumptions of the solar
metallicity BC03 models and the \citet{reddy15} stellar attenuation
curve (Section~\ref{sec:sedmodeling}).  Figure~\ref{fig:ebmvcompare}
shows the comparison between the reddenings for galaxies with $S/N \ge
5$ in both $\ha$ and $\hb$ and where $\ebmvgas \le 1.0$.

A Spearman correlation test indicates that $\ebmvgas$ is correlated
with $\ebmvstars$ with $\simeq 8\sigma$ significance, and linear
regression between the two variables (keeping the intercept fixed at
zero) gives the following relation:
\begin{eqnarray}
\ebmvgas = (2.070\pm 0.088)\times \ebmvstars.
\end{eqnarray}
The trend between $\ebmvgas$ and $\ebmvstars$ is not significantly
affected when accounting for galaxies that have undetected $\hb$ lines
($S/N<3$) based on stacking in bins of $\ebmvstars$
(Figure~\ref{fig:ebmvcompare}).

The data indicate a relationship between nebular and stellar reddening
that is similar in slope to the canonical relation from
\citet{calzetti00}, i.e., $\ebmvgas = 2.27\times \ebmvstars$.
Figure~\ref{fig:ebmvcompare} shows the relations between $\ebmvgas$
and $\ebmvstars$ when assuming other stellar attenuation curves for
deriving $\ebmvstars$.  For example, \citet{shivaei20} find that
MOSDEF galaxies in the upper $\sim$half of the stellar mass
distribution ($M_\ast \ga 10^{10.4}$\,$M_\odot$) have a shallower
(Calzetti-like) stellar attenuation curve, while those in the lower
$\sim$half of the mass distribution have a steeper (SMC-like) curve.
These results are consistent with other studies that have found
steeper curves for high-redshift ($z \ga 2$) galaxies with lower
stellar masses (e.g., \citealt{reddy06b, reddy10, reddy12a,
  bouwens16a, fudamoto17, fudamoto20}).  Based on these previous
studies, we also modeled the galaxies assuming the SMC extinction
curve with the $0.2$\,$Z_\odot$ metallicity models for those with
$\ebmvgas \le 0.4$ (corresponding to $M_\ast \la
10^{10.4}$\,$M_\odot$; \citealt{shivaei20}) and the \citet{calzetti00}
curve with the $Z_\odot$ metallicity models for those with $\ebmvgas >
0.4$.  The resulting trend between $\ebmvgas$ and $\ebmvstars$ is
indicated by the green line in Figure~\ref{fig:ebmvcompare}
(``SMC+Calz'' trend).  Finally, we show the trend obtained if we
assume the SMC extinction curve for all the galaxies in our sample.

The average ratios between the nebular and stellar reddening are
$\langle \ebmvgas / \ebmvstars\rangle = 2.070, 2.273, 2.712,$ and
$4.331$ for the \citet{reddy15}, \citet{calzetti00}, ``SMC+Calz'', and
SMC curves, respectively.  The ratio increases with the steepness of
the assumed stellar attenuation curve because assuming such a curve
results in smaller $\ebmvstars$.  More generally, the specific
relationship between $\ebmvgas$ and $\ebmvstars$ is of limited value
without knowledge of the attenuation curves used to derive the color
excesses (e.g., \citealt{reddy15, theios19, shivaei20}).  Furthermore,
as is evident from Figure~\ref{fig:ebmvcompare}, the relatively large
scatter between $\ebmvgas$ and $\ebmvstars$ (rms$\simeq 0.22$)---in
conjunction with the fact that they are derived assuming a fixed
average dust curve that may not apply on a galaxy-by-galaxy
basis---implies that the average relationship between the two is only
meaningful in the context of large ensembles of galaxies rather than
individual objects \citep{reddy15, theios19, shivaei20}.  At any rate,
in all cases of the assumed attenuation curves, we find that the
reddening of the nebular lines exceeds that of the stellar continuum
on average.

From a physical standpoint, the increased reddening towards the
ionized regions of high-redshift galaxies implies that the youngest
stellar populations are enshrouded by a higher column density and/or
covering fraction of dust.  We return to this issue in
Section~\ref{sec:physicalpicture}.

\subsection{A Physical Context for the Difference between Nebular and
Stellar Reddening}
\label{sec:physicalpicture}

One of the key results of our analysis is that the Balmer decrements
of typical star-forming galaxies at $z\sim 2$ imply nebular color
excesses that are significantly redder than those measured for the
stellar continuum (Figure~\ref{fig:ebmvcompare}).  This conclusion is
reached irrespective of the attenuation curve assumed for either the
nebular regions or the stellar continuum.  This differential reddening
has been noted in several other studies of high-redshift galaxies
(e.g., \citealt{forster09, kashino13, reddy10, price14, reddy15,
  shivaei20}), and is also seen in local star-forming galaxies (e.g.,
\citealt{fanelli88, calzetti97, calzetti00, kreckel13}).  From a
physical standpoint, the discrepancy between the nebular and stellar
reddening has generally been attributed to the redder lines-of-sight
towards the youngest stellar populations in a galaxy (e.g.,
\citealt{fanelli88, calzetti97, calzetti00, kreckel13}).  Here, we
revisit the physical context for the difference in reddening of the
nebular emission lines and stellar continuum in light of recent
improvements in stellar population modeling that account for the
effects of binary stellar evolution.

The simplest interpretation for the difference in the reddening of the
nebular emission lines and stellar continuum in galaxies is based on
considering the crossing times and/or dissipation timescales for
molecular clouds.  Specifically, the main sequence lifetimes of the
very massive single non-rotating stars that dominate the Balmer
emission-line luminosities (spectral type O6 and earlier;
\citealt{leitherer90}) are shorter than the typical molecular cloud
crossing timescale of $\simeq 1-3$\,Myr \citep{calzetti94}.  Thus,
these very massive stars are producing most of the ionizing flux while
they are still embedded in their birth clouds and in regions of higher
dust column density.  This prodigious source of ionizing photons is
lost once these massive stars go supernovae and the molecular cloud is
subsequently disrupted.  Those OB associations that are observed after
their clouds have dissipated, and which lack the most massive O stars,
produce fewer ionizing photons and do not contribute as much to the
Balmer emission lines, while still contributing significantly to the
non-ionizing stellar continuum.  Hence, the Balmer lines primarily
originate from regions around the most massive stars that are subject
to additional reddening beyond that affecting the stellar continuum.

Here, we re-evaluate the picture described above in the context of
recent studies that demonstrate the necessity of binary stellar
populations (or rotating massive stars) to jointly reproduce the
rest-frame UV and rest-frame optical spectra of typical galaxies at
$z\sim 2$.  Specifically, several studies argue that low stellar
metallicity massive binary stellar populations are required to
reproduce the observed strong-line ratios (e.g., $\oiii/\hb$ versus
$\nii/\ha$) of $z\sim 2$ galaxies, while simultaneously matching the
rest-frame far UV spectra (e.g., \citealt{steidel14, steidel16,
  topping20}).  Three consequences of these massive binary star models
are that they increase massive star main sequence lifetimes, result in
a broader range of stellar masses over which ionizing photon
production occurs, and boost the ionizing flux per unit star-formation
rate \citep{eldridge17}.  At subsolar stellar metallicities ($Z\la
0.3Z_\odot$), these models predict an H-ionizing flux that peaks
roughly 3\,Myr---and falls by a factor of $\approx 3$ roughly
10\,Myr---after an instantaneous burst of star formation
\citep{stanway16}.  Such models also produce a factor of $\ga 5\times$
the ionizing flux as single star models 10\,Myr after an initial burst
of star formation.\footnote{For a fixed IMF, metallicity,
  star-formation history, and age, stellar population synthesis models
  that include the effects of binary evolution yield higher ionizing
  fluxes than single star models, but the shape of the non-ionizing UV
  continuum is very similar between the two.  As $\ebmvstars$ is
  determined from the non-ionizing UV continuum, these values are
  largely insensitive to the effects of binary stellar evolution.}

The increase in, and duration of, ionizing flux production predicted
by these models suggest that significant ionizing flux may persist
even after an OB association drifts away from the parent GMC on a
timescale of $\la 3$\,Myr.  Once the massive stars are free from the
obscuring dust associated with their natal clouds, the line emission
from the nebulae surrounding these stars would be subject to
essentially the same columns of dust that affect the stellar
continuum, resulting in a nebular reddening that is more in line with
that measured for the stellar continuum.  The observed differences
between the nebular and stellar reddening suggest that the molecular
cloud crossing timescales may be at least factor of $3\times$ longer
than those of local GMCs, and closer to the typical cloud disruption
timescales of $\simeq 10-30$\,Myr \citep{blitz80, mckee07}.

A second likely possibility for the differential reddening between the
nebular lines and stellar continuum is related to the constant or
rising star-formation histories that are favored for typical
star-forming galaxies at $z\ga 2$ (e.g., \citealt{papovich11,
  reddy12b}).  For these star-formation histories, the lower ionizing
flux of slightly older OB associations that are dissociated from their
birth clouds will be compensated by the higher ionizing flux of
newly-formed and dustier OB associations, where the latter will
contribute significantly to the galaxy-averaged $\ebmvgas$.

A third and equally likely possibility is that the dust
that reddens the Balmer line photons may not be localized to the
molecular clouds, but distributed more widely in the ISM, i.e., in the
ionized and neutral gas phases.  In this case, even after the parent
molecular cloud has dissipated, the nebular line emission may be
dominated by those recently-exposed OB associations that are still
located preferentially in regions of higher dust column density as
averaged over all phases of the ISM.  Indeed, modeling of the far-UV
spectra of $z\sim 3$ galaxies suggests that a significant fraction of
dust is not associated with the molecular phase of the ISM in these
galaxies \citep{reddy16a}.  This conclusion is consistent with the
almost ubiquitous presence of metals (and hence dust) in the neutral
and ionized ISM as indicated by low- and high-ionization interstellar
metal absorption lines in the rest-frame far-UV spectra of
high-redshift galaxies (e.g., \citealt{shapley03}).  

In general, the limited spatial resolution characteristic of
high-redshift galaxy observations implies that global line and
continuum measurements undoubtedly include many OB associations.
Thus, the difference in $\ebmvgas$ and $\ebmvstars$ (or their
similarity for some subsets of high-redshift galaxies; e.g.,
\citealt{reddy10, reddy15, pannella15, shivaei20}) is likely driven by
the distribution of dust column densities along the lines-of-sight to
different OB associations.  Determining which of the above
possibilities, if any, may be most relevant for explaining the degree
of difference in the reddening of the nebular lines and stellar
continuum will require detailed simulations of small-scale
star-formation, the associated feedback, dust formation, and the
subsequent evolution of the GMCs when adopting stellar population
models that include the effects of stellar multiplicity.

\section{\bf CONCLUSIONS}
\label{sec:conclusions}

We use deep rest-frame optical spectra of 532 star-forming galaxies at
redshifts $z\simeq 1.4-2.6$ from the MOSDEF survey to place the first
constraints on the nebular attenuation curve at high redshift.
Specifically, we use the first five low-order Balmer emission lines
($\ha$, $\hb$, $\hg$, $\hd$, and $\he$) detected in the composite
spectra of these galaxies to infer the shape and normalization of the
nebular attenuation curve at rest-frame optical wavelengths.  

The nebular attenuation curve derived here is similar in shape to that
of the Galactic extinction curve at rest-frame optical wavelengths, a
result that is insensitive to the functional form assumed for the
curve (Section~\ref{sec:shapeofcurve}; Figure~\ref{fig:kave}).  The
derived ratio of the total-to-selective absorption at $V$-band depends
on the extrapolation of the nebular attenuation curve to long
wavelengths and lies in the range $R_V \approx 3.1-3.3$.  Within the
systematic uncertainties, these values of $R_V$ are similar to the
$R_V = 3.1$ for the Galactic extinction curve
(Section~\ref{sec:normofcurve}).  If the similarity in shape and
normalization of the attenuation curve extend to rest-frame UV
wavelengths, then our results suggest that the dust ``seen'' along the
nebular sightlines in $z\sim 2$ galaxies can be approximated as a
foreground screen of dust that has similar grain scattering/absorption
properties and a size distribution as those inferred for the average
Galactic sightline.

If we assume the stellar attenuation curve of \citet{reddy15}, we
obtain a relationship between $\ebmvgas$ and $\ebmvstars$ that is
similar in slope to that of the commonly-assumed \citet{calzetti00}
relation (Section~\ref{sec:reddening}).  The exact slope will of course
depend on the choice of stellar attenuation curve: steeper curves
results in higher slopes and higher average ratios of
nebular-to-stellar reddening.  Regardless of the adopted curve for the
stellar continuum, however, we find that the nebular reddening is on
average larger than that of the stellar continuum.

We discuss the physical context for this differential reddening of the
nebular lines and stellar continuum in light of recent results that
favor binary population synthesis models in reproducing the rest-frame
UV and rest-frame optical spectra of high-redshift galaxies.  These
models predict ionizing photon fluxes that are a few times larger, and
can be sustained for longer periods, relative to those
obtained from single star models.  In this framework, the difference in
reddening of the nebular lines and stellar continuum may imply
molecular cloud crossing times in excess of a few Myr, star-formation
rates that are constant or increase with time such that newly-formed
(and dustier) OB associations always dominate the total ionizing flux,
or may indicate that the dust that dominates the reddening of the
Balmer lines is associated with the non-molecular (i.e., neutral and
ionized) phases of the ISM (Section~\ref{sec:physicalpicture}).

The analysis presented here hints at the rich array of studies that
will be enabled with direct measurements of multiple low- and
high-order nebular recombination emission lines for individual
galaxies with the next generation of ground- (e.g., $\ga 30$\,m class)
and space-based (e.g., {\em James Webb Space Telescope}) facilities.
The effective depths of the composite spectra used here to constrain
the nebular attenuation curve have limited our analysis to an
examination of the average shape and normalization of the dust curve
across our entire sample.  Aside from enabling nebular dust
corrections on an object-by-object basis, direct detections of the
higher-order Balmer lines and longer-wavelength Paschen and Brackett
lines, along with constraints from the rest-frame UV (e.g.,
\ion{He}{2} recombination lines), will allow for analyses of how the
normalization and shape of the curve vary from galaxy-to-galaxy, and
how the curve may depend on metallicity, stellar mass, SFR, and other
characteristics.  Such analyses will provide a powerful new probe of
the lifecycle, properties, and spatial distribution of dust in
galaxies.

\acknowledgements

We acknowledge support from NSF AAG grants AST-1312780, 1312547,
1312764, and 1313171; archival grant AR-13907 provided by NASA through
the Space Telescope Science Institute; and grant NNX16AF54G from the
NASA ADAP program.  We are grateful to the MOSFIRE instrument team for
building this powerful instrument, and to Marc Kassis at the Keck
Observatory for his many valuable contributions to the execution of
the MOSDEF survey. We also acknowledge the 3D-HST collaboration, who
provided us with spectroscopic and photometric catalogs used to select
MOSDEF targets and derive stellar population parameters. We also thank
I. McLean, K. Kulas, and G. Mace for taking observations for the
MOSDEF survey in May and June 2013.  We wish to extend special thanks
to those of Hawaiian ancestry on whose sacred mountain we are
privileged to be guests.  Without their generous hospitality, the
observations presented herein would not have been possible.


\end{document}